\documentclass[12pt]{article}
\usepackage{amsmath,citesort,amssymb,mathrsfs,url,graphicx}

\title{The Bohmian Approach to the Problems of Cosmological Quantum Fluctuations}

\author{Sheldon Goldstein\footnote{Department of Mathematics,
     Rutgers University, Hill Center, 
     110 Frelinghuysen Road, Piscataway, NJ 08854-8019, 
     USA. E-mail: oldstein@math.rutgers.edu},
Ward Struyve\footnote{Department of Physics,
	Universit\'e de Li\`ege,
	B\^atiment B15, Sart Tilman,
	4000 Li\`ege, Belgium. E-mail: ward.struyve@ulg.ac.be},
Roderich Tumulka\footnote{Department of Mathematics,
     Rutgers University, Hill Center, 
     110 Frelinghuysen Road, Piscataway, NJ 08854-8019, 
     USA. E-mail: tumulka@math.rutgers.edu}
}

\date{August 5, 2015}

\newcommand{\dd}{\mathrm{d}}

\def\pa{\partial}

\def\al{\alpha}

\def\ii{\mathrm i}
\def\ee{\mathrm e}

\newcommand{\be}{\begin{equation}}
\newcommand{\en}{\end{equation}}

\begin{document}
\maketitle

\begin{abstract}
There are two kinds of quantum fluctuations relevant to cosmology that we
focus on in this article: those that form the seeds for structure
formation in the early universe and those giving rise to Boltzmann brains
in the late universe. First, structure formation requires slight
inhomogeneities in the density of matter in the early universe, which
then get amplified by the effect of gravity, leading to clumping of
matter into stars and galaxies. According to inflation theory, quantum
fluctuations form the seeds of these inhomogeneities. However, these
quantum fluctuations are described by a quantum state which is
homogeneous and isotropic, and this raises a problem, connected to the
foundations of quantum theory, as the unitary evolution alone cannot
break the symmetry of the quantum state. Second, Boltzmann brains are
random agglomerates of particles that, by extreme coincidence, form
functioning brains. Unlikely as these coincidences are, they seem to be
predicted to occur in a quantum universe as vacuum fluctuations if the
universe continues to exist for an infinite (or just very long) time, in
fact to occur over and over, even forming the majority of all brains in
the history of the universe. We provide a brief introduction to the
Bohmian version of quantum theory and explain why in this version,
Boltzmann brains, an undesirable kind of fluctuation, do not occur (or at
least not often), while inhomogeneous seeds for structure formation, a
desirable kind of fluctuation, do.
\end{abstract}
\bibliographystyle{unsrt}

\section{Introduction}
The notion of observation or measurement plays a fundamental role in standard
quantum theory. One of the basic postulates of quantum mechanics is namely that
upon measurement the Schr\"odinger evolution of the wave function is interrupted
by a collapse. However, the notions of observer or measurement are
rather vague. Exactly which physical processes  count as measurements? What
counts as an observer? All living creatures? Or only humans? 
Or only humans with a Ph.D.? And why should there be special rules for
measurement in quantum mechanics to begin with, without which there would
apparently be no collapses---and no results of measurements? This is one version of the
measurement problem of quantum mechanics.

This problem is especially severe in the context of cosmology. How can we
describe the early universe when there were no measurements or observers? An
adequate description requires what Bell \cite{bell86,bell87a} dubbed a {\em quantum theory without observers}, a theory in which observers or measurements do not play a fundamental role. One such theory is Bohmian mechanics \cite{bohm52ab,goldstein06}. Bohmian mechanics describes a configuration of particles and/or fields and/or metrics, or in general some kind of variables associated with space-time points, called \emph{local be-ables} (as opposed to observ-ables), which evolve under the influence of the wave function. In Bohmian mechanics it is the local beables and their behavior, not measurement or observation, that is fundamental. 

In this paper we want to consider two problems in quantum cosmology which
are closely related to the quantum measurement problem and explain how they get resolved in the context of
Bohmian mechanics. These problems have to do with vacuum fluctuations. On the
one hand, according to current cosmological theories, all structure (i.e., non-uniformity of matter, such as stars and galaxies) is supposed
to come out of vacuum fluctuations. The problem, emphasized particularly by Sudarsky and collaborators \cite{perez06,sudarsky11,okon14}, is how these quantum fluctuations can 
turn into classical perturbations of the
matter density. The other problem, emphasized by Boddy, Carroll, and Polack \cite{boddy14,boddy15}, 
has to do with vacuum fluctuations at late
times. At late times, it is expected that the universe will be driven to homogeneity 
due to the continuing expansion of the universe, driven by the cosmological constant.
All that remains
are vacuum fluctuations. These vacuum fluctuations are believed to yield every
possible configuration over time, in particular ``Boltzmann brains'' which, if the
universe continues to exist long enough, will outnumber the ordinary
brains. According to the ``Copernican principle'' \cite{gott93}, 
which states that we should expect not to occupy a privileged position in the universe, we should expect to be one of the Boltzmann brains. 
But our observations show clearly that we are not Boltzmann brains, which puts our cosmological theory into question.

The outline of the paper is as follows. The problems of cosmological quantum fluctuations are described in more detail in section~\ref{fluctuations}. In section~\ref{bm}, we provide an introduction to Bohmian mechanics and explain the basic mechanisms with which the Bohmian approach can address these problems. In section~\ref{flrw}, we describe a concrete Bohmian model of quantum field theory on a curved background space-time with an expanding Friedman-Lema\^itre-Robertson-Walker metric, in terms of which we analyze the problem of Boltzmann brains and, in section~\ref{bohmfluctuations}, the problem of classical perturbations from quantum fluctuations.

\section{Problems of vacuum fluctuations in cosmology}\label{fluctuations}

According to current cosmological models, our universe is believed to spatially
expand exponentially fast both at early and late times. According to inflation
theory, which describes the early times, the inflaton field is driving the
expansion. At late times, the cosmological constant is causing the expansion. 
In both cases, the expansion drives the universe to homogeneity,
so that nothing remains but vacuum fluctuations. These
vacuum fluctuations are
regarded as a virtue in the early universe, but as a vice in the late universe.
In the early universe they are considered to be the seeds of structure
formation, while in the late universe they lead to the problem of Boltzmann
brains.

\subsection{Problem with seeds of structure formation}
\label{sec:problemseeds}

To illustrate the nature of the difficulty, it may be helpful to consider
the following simple example. As a toy model, consider non-relativistic,
Newtonian gravity for $N$ point particles in a large box $\Lambda=[0,L]^3$ with
periodic boundary conditions. According to classical mechanics, if this
system starts out in a near-uniform configuration (with initial
velocities, say, equal to zero, or random with a Gaussian distribution),
it evolves to a clumped configuration under the influence of gravity; that
is, gravity will amplify any slight initial non-uniformity. Now in quantum
mechanics, consider a constant wave function $\psi_0$ on configuration
space $\Lambda^N$; it gives $>99\%$ weight to near-uniform configurations
and evolves, under the Schr\"odinger equation over a suitable duration $t$,
to a wave function $\psi_t$ that gives $>99\%$ weight to clumped
configurations but is still invariant under 3-translations because
$\psi_0$ and the Hamiltonian are. So $\psi_t$ has a symmetry that the clumped
configuration we observe does not have.
That is, $\psi_t$ is a superposition of many ``clumped states,'' and not a
random clumped state, and
$\psi_t$ contains no information about which clumped configuration is the
real one. This problem is, of course, a variant of the usual measurement
problem of quantum mechanics, but it is exacerbated in cosmology, where we
cannot exploit what an observer outside the system would see upon a
measurement.

The same kind of problem arises in current cosmological theories, such as inflation theory. 
The formation of structure such as stars and galaxies requires initial perturbations in the
matter
density. According to inflation theory, the vacuum fluctuations of the metric
field and inflaton field are the seeds for these matter density perturbations. 
During the inflationary expansion, the
inflaton and metric fields are approximately homogeneous and isotropic.
The deviations from homogeneity and isotropy are described by the vacuum state which, however, is itself homogeneous and isotropic as a quantum state.
This inflationary period lasts until the inflaton field decays into ordinary
matter fields. It is
believed that during the inflationary period these quantum vacuum fluctuations became
classical field perturbations. Once the inflaton field decays into ordinary
matter, the matter distribution inherits the non-homogeneity of these perturbations. 
The inhomogeneities in the matter density then grow because of
gravitational clumping, to eventually form large scale structures such as stars,
galaxies and clusters of galaxies. An imprint of these fluctuations at the time of
decoupling (when the photons were allowed to move freely) can be found in
the temperature pattern of the micro-wave background radiation. The precise
details of these temperature fluctuations allows us to distinguish between different
inflationary models. 

This is regarded as part of the success story of inflation theory
\cite{liddle00,mukhanov05,weinberg08,lyth09,peter09}. However, there is still a
problem with the standard presentation. Namely, how exactly did the vacuum
fluctuations, which are described by a quantum state that is homogeneous and
isotropic, give rise to classical field perturbations that no longer have
this symmetry? After all, the unitary evolution preserves the symmetry.
Hence, according to standard quantum theory, the symmetry can only be broken by
collapse of the wave function. But collapse of the wave function is supposed to
happen upon measurement. What physical process is playing the role of a
measurement in the early universe? There are no measurement devices or
observers around at that time. Moreover, objects 
such as measurement devices or observers (which are
obviously inhomogeneous) are themselves supposed to stem, ultimately, from the
quantum fluctuations in the metric and inflaton fields. 

In order to address this problem, one needs a precise version of
quantum mechanics, which does not suffer from the measurement problem. One class
of precise approaches is collapse theories, where collapses happen spontaneously
at random times, according to an observer-independent law. An analysis of
the problem at hand in the context of collapse theories has been carried out by a
number of people, especially by Sudarsky and collaborators
\cite{perez06,deunanue08,leon10,sudarsky11,martin12,canate13,okon14}. An 
analysis in
terms of Bohmian mechanics has also been carried out
\cite{hiley95,pinto-neto12a}. We will
turn to it in sections \ref{sec:breaking} and \ref{bohmfluctuations}.

\subsection{Problem with Boltzmann brains}

In addition, at late times, there is the worry of Boltzmann brains \cite{albrecht01,dyson02,albrecht04,bousso07,boddy14,boddy15}. 

We begin by explaining what a Boltzmann brain is. Let $M$ be the present
macro-state of your brain. 
For a classical gas in thermal equilibrium, after sufficient waiting time some atoms will, with  probability 1, ``by coincidence'' (or ``by fluctuation'') happen to come together in such a way as to form a subsystem in a micro-state compatible with $M$ (at least if the classical dynamics is ``sufficiently ergodic,'' which we shall assume here). 
That is, this brain
comes into existence not by fetal development and, prior to that, evolution of life forms, but by
coincidence; this brain has memories (duplicates of your present
memories), but they are false memories: the events described in the
memories never happened to this brain. Boltzmann brains are, of course,
\emph{very} unlikely. But they will form if the waiting time is long
enough, and they will form more frequently if the system is larger (larger
volume and number of particles).

Let us explain why Boltzmann brains can lead to a problem. According to the ``Copernican principle'' \cite{gott93}, we should be typical observers in a typical universe. This principle can be
understood as a rule for extracting predictions from a theory: we should see what a typical observer (weighted with life span) in a typical universe sees. 
For example, consider classical mechanics with
the additional constraint that the initial micro-state of the universe is
compatible with a certain low-entropy macro-state; in a typical solution
of the dynamical equations (``typical universe''), entropy increases with
time, so, according to the Copernican principle, the theory predicts that
we should see entropy increase; this is, in fact, Boltzmann's explanation
of the observed entropy increase. 

Now the following problem with Boltzmann brains arises.
Imagine for example a classical non-relativistic universe in a finite volume. Suppose that the universe continues to exist forever, so that the whole universe reaches thermal equilibrium at some time in the distant future. After an extremely long waiting time, a Boltzmann brain will spontaneously form out of some particles in the thermal equilibrium; since the universe exists forever, this unlikely event is certain to happen sooner or later. After a short while the Boltzmann brain will disintegrate, and the universe will again be in thermal equilibrium. After another extremely long waiting time, the next Boltzmann brain will form, probably in another place and from other particles. If time is infinite, this is again certain to happen. And to happen again and again. In fact, the number of times this will happen is by far greater than the total number of brains that ever grew as part of the evolution of life before the universe reached global thermal equilibrium. Thus, the overwhelming majority of
brains in the universe will be Boltzmann brains. According to the
Copernican principle, the theory predicts that we are Boltzmann brains.
But we are not. After all, as emphasized by Feynman \cite[p.~115]{feynman95}, most fluctuations leading to Boltzmann brains will be no larger than necessary, so Boltzmann brains find themselves surrounded by thermal equilibrium, not by other living beings on a tepid planet orbiting a hot star. So, the theory makes a wrong prediction. The question is
how any of our serious theories can avoid making this wrong prediction.

Here is a concrete version of the problem. As emphasized by Boddy, Carroll, and Polack
\cite{boddy14,boddy15}, there is 
reason to believe that the late universe will be
close to de Sitter space-time (e.g., this occurs in the $\Lambda$CDM model), 
that all matter will be driven to homogeneity, and that
the quantum state that describes the deviations from uniformity
will locally look like the Bunch-Davies vacuum, 
a quantum state that is invariant under all isometries of de Sitter space-time. 
For the sake of definiteness, let us take for granted in the following that the late universe is approximately de Sitter, at least in large regions, although the discussion of the problem does not depend much on that. 
On any kind of configuration space (be it of particle or field configurations), 
the probability distribution that the Bunch-Davies vacuum defines 
gives positive probability to all sorts of configurations, including Boltzmann brain configurations. 
Does this mean there are Boltzmann brains sparsely scattered around in the Bunch-Davies vacuum? Or, could Boltzmann brains occur, rarely but over and over during an infinite amount of time, in the Bunch-Davies vacuum?

This leads to the question, What is the significance of this particular
wave function for reality? Does a stationary state mean that nothing
happens? Carroll adheres to the many-worlds view of quantum mechanics, and
he and other advocates of many-worlds (such as David Wallace \cite{wallace12}) 
disagree about whether a stationary state means that nothing happens. If it means
that nothing happens, this would remove the problem, as we discuss in section~\ref{sec:freezingBoltzmann} below. 
While the many-worlds view seems
ambiguous about this question, the Bohmian approach naturally provides
an unambiguous answer: in any given Bohm-type model, it is clear whether Boltzmann brains
do or do not occur. We will argue in sections
\ref{sec:freezingBoltzmann} and \ref{sec:noboltz} that in natural models, they do not occur.

\section{Bohmian mechanics}\label{bm}

We now describe non-relativistic Bohmian mechanics (also called pilot-wave theory or
de Broglie-Bohm theory), a theory about point particles in physical
space moving under the influence of the wave function. Later in section~\ref{flrw}, we describe an extension of this theory for a field in curved space-time. 

\subsection{Definition of the theory}

\begin{figure}[h]
\begin{center}
\includegraphics[width=10cm]{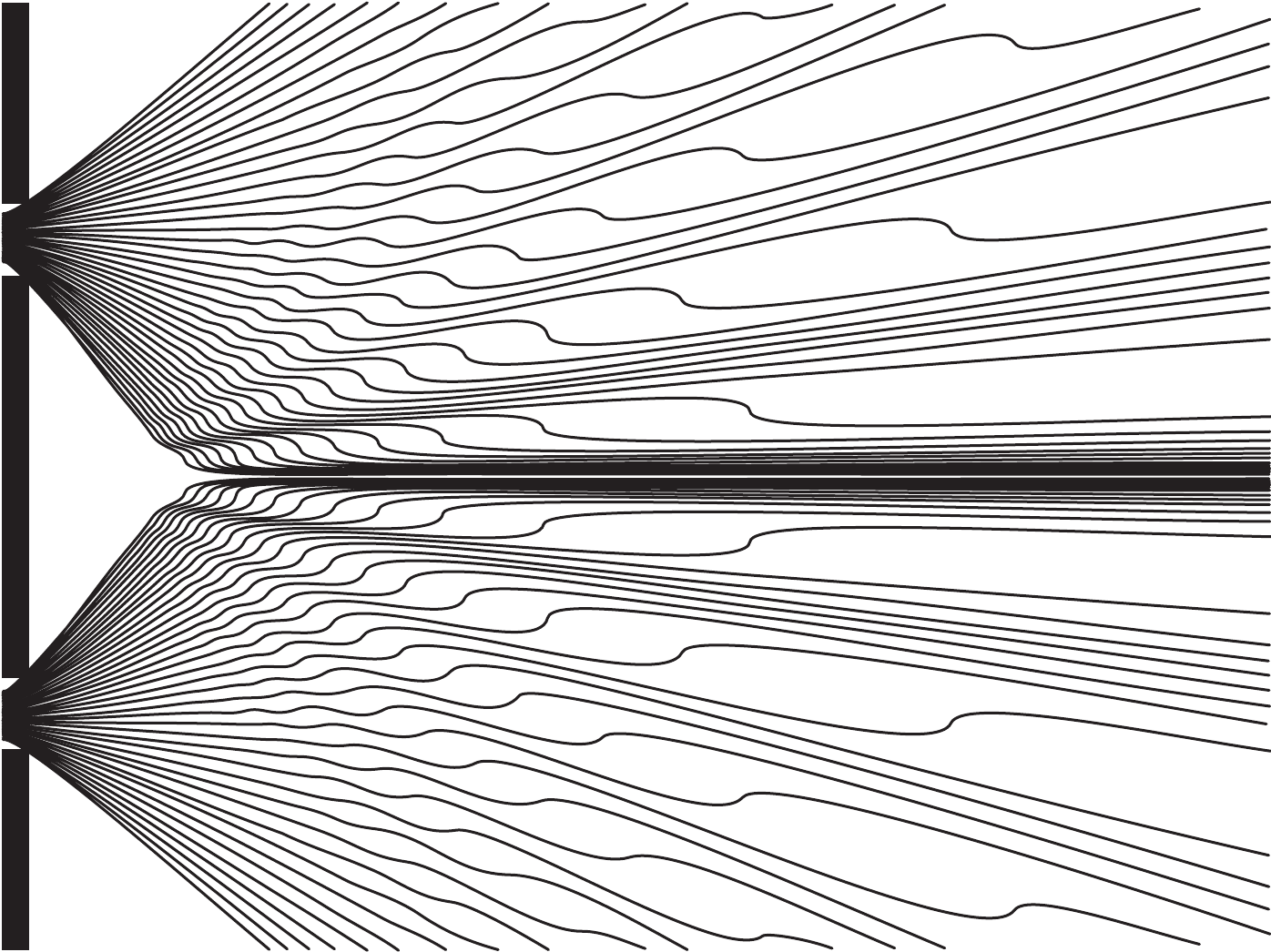}
\end{center}
\caption{Several alternative possible Bohmian trajectories of a particle passing through a double-slit setup with the wave function passing through both slits. The selection of trajectories shown follows roughly the $|\psi|^2$ distribution. Picture created by G.~Bauer after \cite{philippidis79}.}
\label{fig:2slit}
\end{figure}

The equation of motion
for the configuration $X=({\bf X}_1,\dots,{\bf X}_n)$ of the particle positions
reads\footnote{Throughout the paper we assume units in which $\hbar=c=1$.}
\be
{\dot X}(t) = v^\psi(X(t),t)
\label{1}
\en
where the velocity vector field $v^\psi=({\bf v}^\psi_1, \dots , {\bf v}^\psi_n)$ is given by
\be
{\bf v}^\psi_k = \frac{1}{m} {\textrm{Im}}\left(
\frac{\psi^*\boldsymbol{\nabla}_k \psi}{\psi^*\psi} \right) \,.
\label{2a}
\en
For particles without spin this assumes the form
\be
{\bf v}^\psi_k = \frac{1}{m} {\textrm{Im}}\left(
\frac{\boldsymbol{\nabla}_k \psi}{\psi} \right) =  \frac{1}{m}
\boldsymbol{\nabla}_k S \,, \qquad \psi = |\psi| \ee^{\ii S} \,.
\label{2b}
\en
The wave function $\psi(x,t)=\psi({\bf x}_1,\dots,{\bf x}_n,t)$ itself satisfies
the non-relativistic Schr\"odinger equation
\be
\ii \pa_t \psi(x,t) = \left( - \sum^n_{k=1} \frac{1}{2m_k} \boldsymbol{\nabla}^2_k + V(x)
\right) \psi(x,t) \,.
\label{3}
\en
Examples of Bohmian trajectories are shown in Figure~\ref{fig:2slit} for a single particle in a double-slit setup.

For an ensemble of systems all with the same wave function $\psi$, there is a
distinguished distribution given by $|\psi|^2$, which is called the {\em quantum
equilibrium distribution}. This distribution is {\em equivariant}:
it is preserved by the particles dynamics \eqref{1} in the sense that if the
particle distribution is given by $|\psi(x,t_0)|^2$ at some time $t_0$, then it
is given by $|\psi(x,t)|^2$ at all times $t$. This follows from the facts that
any distribution $\rho$ that is transported by the particle motion satisfies the
continuity equation
\be
\pa_t \rho + \sum^n_{k=1} \boldsymbol{\nabla}_k\cdot (\rho {\bf v}^\psi_k) = 0 
\label{4}
\en
and that $|\psi|^2$ satisfies the same equation, i.e.,
\be
\pa_t |\psi|^2 + \sum^n_{k=1} \boldsymbol{\nabla}_k\cdot(|\psi|^2 {\bf v}^\psi_k) = 0 
\label{5}
\en
as a consequence of the Schr\"odinger equation \eqref{3}. 

We note that, as a consequence of the equation of motion \eqref{1} with \eqref{2a} or \eqref{2b}, if the wave function factorizes, as in
\be
\psi({\bf x}_1,\ldots,{\bf x}_n)=\varphi({\bf x}_1,\ldots,{\bf x}_m) \, \chi({\bf x}_{m+1},\ldots, {\bf x}_n)\,,
\en
then the subsystem ${\bf X}_1,\ldots,{\bf X}_m$ obeys the analogous equation of motion involving only $\varphi$, as $\chi$ cancels out.

In Bohmian mechanics, it is meaningful to speak of the wave function of the universe because, in the Bohmian approach, the wave function is not primarily regarded as a tool for predicting what observers will see but as a part of the physical world. In particular, Equations \eqref{1}--\eqref{3} define a simple, non-relativistic model of the universe with $n$ particles, with $\psi$ the wave function of the universe. Assuming (as we will henceforth) that the initial configuration of the universe $X(t=0)$ is typical with respect to the $|\psi(x,0)|^2$ distribution (i.e., as if randomly chosen with $|\psi(x,0)|^2$ distribution), it can be shown \cite{duerr92a,duerr12,duerr04a} that the empirical predictions of Bohmian mechanics agree with the standard quantum formalism.

Note that the velocity vector field is of the form $v^\psi=j^\psi/|\psi|^2$, where
$j^\psi=({\bf j}^\psi_1,\dots,{\bf j}^\psi_n)$ with ${\bf j}^\psi_k= 
{\textrm{Im}}( \psi^* \boldsymbol{\nabla}_k \psi )/m $ is the usual quantum-mechanical probability current. In other quantum theories, such as for example quantum field theories,
the equation of motion can be set up in a similar way by dividing the appropriate current
by the density. In this way it is ensured that the dynamics will leave the
density equivariant. (See for example \cite{struyve09a} for a treatment of
arbitrary Hamiltonians.)

In a Bohmian universe, there is the wave function $\psi$ of the universe and
a configuration $Q$ of the universe describing the arrangement of all
fundamental fields and particles. There is also a natural definition for the
wave function $\varphi$ of a subsystem with configuration
$X$---corresponding to a splitting $Q=(X,Y)$ of the configuration of the
universe into that of the subsystem and that of its environment, $Y$---called
the \emph{conditional wave function}. For example in the context of non-relativistic
quantum mechanics, the conditional wave function of a subsystem is given by
$\varphi(x,t) = \psi(x,Y(t),t)$. An analysis shows that the conditional wave function collapses according to the
usual text book rules  when a quantum 
measurement is performed \cite{duerr92a,duerr04a,goldstein10}, whereas the wave function $\psi$ of the universe never collapses but evolves unitarily.

\subsection{Extensions of Bohmian mechanics to quantum field theory and quantum gravity}

Non-relativistic Bohmian mechanics has been extended to quantum field theory
\cite{duerr04b,duerr05a,colin07,struyve10,struyve11} and to quantum gravity
\cite{shtanov96,goldstein04,pinto-neto05}. Some approaches \cite{duerr04b,duerr05a,colin07} use a particle ontology (i.e., actual particles with world lines as in \eqref{1} above), others \cite{struyve10,struyve11} a field ontology (i.e., a field configuration guided by the state vector as in \eqref{fieldguidance} below), and others a combination, such as particles for fermions and fields for bosons. The cited
extensions to quantum gravity involve a quantum state that is a wave function on
the space of 3-metrics, governed by the Wheeler-DeWitt equation, along with an actual 3-metric---just as non-relativistic Bohmian mechanics involves $\psi$, a wave function on
the space of particle configurations, 
along with an actual particle configuration $X$. Rather than expressing a time evolution, the
Wheeler-DeWitt equation concerns a \emph{time-independent} wave function, on which it expresses a constraint.
This leads to the problem of time in quantum gravity:
If the wave function must be static, how can we account for the apparent change
of things over the course of time? In particular, how can we account
for the expansion of the universe in terms of a static wave function? This
problem is trivially solved in the Bohmian approach. While the wave function may
be static, the actual 3-metric, fields, and particles generically will not be. Whether or not the
universe is expanding depends on the evolution of this 3-metric.

\subsection{Bohmian mechanics and the breaking of symmetry}
\label{sec:breaking}

The problem of structure seeds is how to obtain, from a wave function with a symmetry, a situation that does not have this symmetry. This problem turns out to be absent in the Bohmian approach because there are further variables, i.e., the local beables, describing a configuration of particles, fields, or other things, besides the wave function. So even when the wave function has a symmetry, the local beables need not share this symmetry. 

The simplest example of this kind is perhaps the double-slit experiment: the wave function passes through both slits (so it is symmetric relative to the axis of the double-slit arrangement) whereas the Bohmian particle passes through only one slit (see Figure~\ref{fig:2slit}), thereby breaking the symmetry. 

Likewise, in the example described at the beginning of section~\ref{sec:problemseeds}, involving $N$ non-relativistic particles with Newtonian gravity in a box $\Lambda=[0,L]^3$ with periodic boundary conditions, the wave function $\psi_t$ is invariant under 3-translations at all times, but the actual configuration $X(t)$ is never invariant under 3-translations, as it consists of $N$ points, and shifting the $N$ points will lead to a different configuration (except for very special translation vectors). So the actual configuration $X$ provides exactly the inhomogeneous, clumped situation that was missing when we considered only the $\psi_t$ obtained through the unitary evolution.

Another example is the symmetry breaking for a decaying atom. Consider a nucleus emitting an $\al$ particle with a spherical wave function. The problem of accounting for the straight tracks
thereby produced  in a cloud chamber was raised by Mott \cite{mott29} in 1929. Suppose for
definiteness that the wave function of the $\alpha$ particle is given by
\be
\psi({\bf x},t) = \frac{\ee^{\ii kr -\ii k^2 t/2m}}{r} \quad \text{with }k>0\text{ and }r=|{\bf x}|\,,
\en
which has spherical symmetry. In the Bohmian description, there is an actual
particle, having an actual position ${\bf X}$ at all times. Its possible
trajectories are given by
\be\label{alpharay}
{\bf X}(t) =  {\bf X}_0 + kt \frac{{\bf X}_0}{|{\bf X}_0|}\,,
\en
so the particle moves radially away from the origin. Even though the wave function is spherically symmetric, the particle's trajectory is not---the particle is moving in one
particular direction. (It happens in this case that the trajectory is a classical trajectory, i.e., a solution of Newton's equation of motion; in general, this is not the case for Bohmian trajectories, as illustrated by Figure~\ref{fig:2slit}. It is, however, significant that in some cases the Bohmian trajectory is classical, even without measurements or decoherence arising from coupling to other degrees of freedom, particularly because in inflation theory one wants to argue that the fluctuations behave classically; we will return to this point in section~\ref{sec:Bohminflation}.)

If we take the interaction of the $\alpha$ particle with the atoms in the cloud chamber into account, then the wave function $\psi({\bf x}_1,\ldots,{\bf x}_n)$ (with ${\bf x}_1$ the position coordinate of the $\alpha$ particle and ${\bf x}_2,\ldots,{\bf x}_n$ those of all other particles forming the cloud chamber) evolves into a superposition of many contributions which differ in where droplets have formed. All contributions agree in that the droplets form along a straight line, but they differ in where that line lies. Now we run into the quantum measurement problem if we insist on the following assumptions: 

\begin{itemize}
\item[(i)] There are no further variables besides the wave function $\psi$. (This assumption is dropped in Bohmian mechanics, where $X(t)$ is a further variable.) 
\item[(ii)] The wave function $\psi$ evolves unitarily. (This assumption is dropped in collapse theories \cite{bell87c,ghirardi11}, where $\psi$ can ``collapse'' also in the absence of an observer.) 
\item[(iii)] The experiment has a single outcome, i.e., an actual track of droplets. (This assumption is dropped in the many-worlds view.) 
\end{itemize}

If all three assumptions are made, then we run into a difficulty because, according to (ii), $\psi$ evolves into a superposition of droplet locations, rather than a random droplet location, and according to (i) there are no further variables that could represent the locations where the actual droplets formed, demanded by (iii). So, one of (i), (ii), (iii) must be dropped. (After doing that, the quantum measurement problem is solved, although other problems may remain, particularly with the many-worlds view \cite{kent10,allori11}.) In the Bohmian approach, an analysis shows that the droplets form along the trajectory \eqref{alpharay} of the Bohmian $\alpha$ particle.

The upshot of these examples is that the Bohmian approach has the resources to avoid the problem with the seeds of structure formation. We will see in section~\ref{bohmfluctuations} below that indeed the problem is absent in a detailed Bohmian model of quantum field theory in curved space-time.

\subsection{Frozen systems in Bohmian mechanics}
\label{sec:frozen}

The phenomenon of frozen systems is a feature of the Bohmian approach that plays a role for the question of Boltzmann brains. To explain this, we begin with an example. For a hydrogen atom in the ground state, Bohm's equation of motion \eqref{1} implies that the electron is at rest relative to the proton, since the ground state wave function
\be\label{hydrogen}
\psi_0({\bf x}) = C\ee^{-\kappa |{\bf x}|}
\en
(with suitable constants $C,\kappa>0$, neglecting spin for simplicity) is real, so that the imaginary part in \eqref{2b} vanishes. This is an example of a frozen system.

It may be surprising that in a theory that takes the notion of particle literally and attributes a trajectory to the electron, the electron is neither circling around the nucleus nor jumping around wildly in a way that would lead to a $|\psi_0|^2$ distribution over time. Rather, the position of the electron relative to the nucleus is initially $|\psi_0|^2$ distributed and remains so due to the lack of relative motion. 

More generally, for Bohmian mechanics with any number of particles and any potential $V$ in the Schr\"odinger equation \eqref{3}, every non-degenerate eigenstate $\psi_{\ell}$ of the Hamiltonian has vanishing Bohmian velocity vector field $v^{\psi_\ell}$ as in \eqref{2a}. 
This follows from the fact that the Hamiltonian in \eqref{3} is real, so that the complex conjugate $\psi_\ell^*$ of $\psi_\ell$ is also an eigenstate, and hence, since $\psi_\ell$ is non-degenerate, agrees with it up to an irrelevant global phase factor, and hence $\psi_\ell$ can be taken to be real. The same phenomenon of freezing occurs in Bohmian models of quantum field theory using either the field ontology (the field is frozen in a random configuration \cite{struyve10}) or the particle ontology (the velocities \cite{duerr04b,duerr05a,colin07} and, when applicable \cite{duerr04b,duerr05a}, the rates of particle creation and annihilation vanish), whenever the quantum state is a non-degenerate eigenstate of the (reversible) Hamiltonian.

The phenomenon of freezing is counter-intuitive in that the momentum distribution defined by the hydrogen ground state $\psi_0$ is (as for every square-integrable wave function) not concentrated on the origin of momentum space. That is, typical momentum values are nonzero, and therefore one often pictures $\psi_0$ as a situation in which the electron undergoes a certain irreducible motion. However, in Bohmian mechanics the momentum variable ${\bf k}$ (the variable in the Fourier transform $\widehat{\psi}({\bf k})$ of $\psi({\bf x})$) is proportional, \emph{not} to the instantaneous velocity $\dot{\bf X}(t)$, \emph{but} to the asymptotic velocity
\be\label{Xinfty}
\dot{\bf X}(\infty)= \lim_{t\to\infty}\dot{\bf X}(t) = \lim_{t\to\infty} \frac{{\bf X}(t)-{\bf X}(0)}{t}
\en 
that the particle would reach if the potential were turned off. So nonzero momentum does not imply, in Bohmian mechanics, nonzero instantaneous velocity.

\subsection{Freezing and Boltzmann brains}
\label{sec:freezingBoltzmann}

The phenomenon of freezing is a basic mechanism relevant to the problem of Boltzmann brains. As a simplified example, suppose the late universe as a whole were governed by non-relativistic Bohmian mechanics, and its wave function were a non-degenerate eigenstate $\psi_\ell$, so that the configuration of the whole universe would be frozen. 
Then the Boltzmann brain problem would be absent, mainly because in a frozen universe Boltzmann brains cannot spontaneously arise as fluctuations. However there are some subtleties that we wish to consider.  Indeed, there are two views on exactly which kind of situation gives rise to a Boltzmann brain problem, and both views agree that the kind of frozen Bohmian universe just described does not have  one. 

The first, ``optimistic'' view insists that a Boltzmann brain will be problematical only if it is a functioning brain, at least for a short time. In a frozen universe, even if a Boltzmann brain configuration existed, it would not be functioning, exactly because it is frozen. A functioning brain requires in Bohmian mechanics the right kind of configuration \emph{and} the right kind of wave function (just as it would require in classical mechanics the right kind of configuration \emph{and} the right kind of momenta), and while the right kind of configuration may occur, the wave function $\psi_\ell$ is not of the right kind. 

The second, ``pessimistic'' view retains the worry that a mere brain configuration may be problematical as it may encode all memories of the brain and perhaps the present thoughts. If that is so, then it becomes important that in a frozen universe, Boltzmann brain configurations cannot occur over and over (unlike in a classical gas in a box which, due to the eternal irregular motion, repeatedly assumes every configuration over time). To be sure, if a Boltzmann brain configuration occurs once, then it stays forever and has the majority of observer-time since the normal brains are finite in number and lifetime. Thus, in the pessimistic view, this configuration poses a threat to the theory. However, if 3-space is not extremely large, then the $|\psi_\ell|^2$ probability that a Boltzmann brain configuration occurs anywhere in space is tiny. Thus, with overwhelming probability, the problem is absent if 3-space is not extremely large.

The situation we have considered so far in this section (a non-relativistic universe in an energy eigenstate) is not actually the situation in the late universe, especially because, due to the expansion of the universe, the Hamiltonian of the particles or fields changes with time, so eigenstates do not remain eigenstates. However, the Bunch-Davies vacuum, often regarded as the asymptotic quantum state for late times, is invariant under isometries of de Sitter space-time and thus stationary in the appropriate sense for de Sitter space-time. And indeed, as we will see in section~\ref{sec:flrwfreezing} below, freezing of the Bohmian configuration occurs at late times in the Bunch-Davies state and, as shown in \cite{tumulka15}, in a generic quantum state. That is, the Bohmian configuration is frozen at late times but not throughout the entire history. We conclude that the Boltzmann brain problem is absent, at least if 3-space is not extremely large.

What if 3-space \emph{is} extremely large? According to the pessimistic view described above, the theory would make incorrect empirical predictions; put differently, we would have to conclude from the fact that we are not Boltzmann brains that 3-space is not extremely large. According to the optimistic view, in contrast, no problem with Boltzmann brains would arise even for extremely large 3-space, for two reasons. First, freezing at late times entails that there are no functioning Boltzmann brains at late times. Second, how about not-so-late times (such as our present era)? While functioning Boltzmann brains would have substantial probability to occur somewhere in 3-space, they would be outnumbered by normal brains. Let us explain. Normal brains (and their present state) come into existence not by fluctuation but through familiar biological processes, preceded by evolution of life, preceded by formation of stars and galaxies, preceded by the low entropy initial state of the universe. The ultimate difference between a normal brain and a Boltzmann brain is whether the brain arose from the low entropy initial state of the universe or as a fluctuation. In view of our existence, it seems reasonable to believe that intelligent life forms evolve in most galaxies. Given that galaxies form at the same rate in every Hubble volume, one would expect a huge number of normal brains within every Hubble volume, whereas functioning Boltzmann brains should be so rare that any given Hubble volume is unlikely to contain any. Thus, at not-so-late times, normal brains should outnumber the Boltzmann brains by far. (And if 3-space is infinite, so that both normal and Boltzmann brains occur infinitely often, then the density of normal brains should be far greater than that of Boltzmann brains, so the Copernican principle would yield that we should be normal brains.)

\subsection{Equal-time and multi-time fluctuations}

How can a frozen Bohmian system, such as a hydrogen atom in the ground state $\psi_0$ as in \eqref{hydrogen}, be compatible with the prediction of the standard formalism that if we measure the position of the electron repeatedly at different times, the empirical distribution of the results will be $|\psi_0|^2$? That is because the position measurement involves an interaction between the electron and an apparatus, and as a consequence of that, the electron is no longer frozen. In fact, this interaction, together with the process of putting the atom back into the ground state, can be shown \cite{duerr92a} to affect the electron in Bohmian mechanics in such a way that its position in the second round of the experiment is different from, and in fact \emph{independent of}, its measured position in the first round. This is in agreement with the fact mentioned already that the empirical predictions of Bohmian mechanics agree with the standard rules of quantum mechanics. So the interaction with a measuring device has a considerable effect on the particle: In the absence of such a device, the positions at different times will be equal, but the empirical distribution for the results of measurements
performed on the particle would, in a Bohmian universe, be nonetheless given by
the usual  quantum probabilities.

As a side remark, it might be worthwhile to contrast the
cases of equal-time experiments (in which $N$ systems are prepared with the same wave function $\varphi$ at the same time $t$, and the same experiment is carried out on each system) and multi-time experiments (in which a single system is prepared at each of $N$ times $t_1,\ldots, t_N$ with wave function $\varphi$, and the same experiment is carried out at each of these times). For both cases the usual quantum probability rules apply: For an
ensemble of similar experiments on systems always beginning with the same wave
function $\varphi$, the statistics for the results  of these experiments will be
given by the quantum probability distributions associated with $\varphi$,
regardless of whether the systems involved are different quantum systems at the
same time or the same quantum system at different times.
However the Bohmian analyses \cite{duerr92a} required for the two cases are
quite different. The multi-time case requires a delicate analysis, one for
which the fluctuations corresponding to the Born rule and to quantum
probabilities  occur only as a result of measurements performed on the
system whose fluctuations we are concerned with. This is quite different from
the situation with equal-time fluctuations, for which the statistics for the
configurations of the ensemble of systems will be given by the usual quantum
probabilities  regardless of whether or not these
configurations are measured. 

Since a Boltzmann brain is very unlikely to occur anywhere in a Hubble volume if the probability distribution over configuration or phase space is similar to that of thermal equilibrium, the reason Boltzmann brains are likely to occur sooner or later in a classical gas in a finite volume is that, due to ergodic properties of the motion, the phase points $X_{t_1},X_{t_2},\ldots$ after long  (and, say, random) waiting times $t_{n+1}-t_n$ can be regarded as approximately independent random variables. In a Bohmian universe, if the configurations $X_{t_1},X_{t_2},\ldots$ at successive times were independent (say, $|\psi_\ell|^2$-distributed) random variables, then Boltzmann brain configurations would be certain to arise sooner or later (which would be a problem at least according to the ``pessimistic'' view). However, they would be independent only if measurements (say by an external observer) were made. In the absence of measurements, freezing ensures that $X_{t_1}=X_{t_2}=\ldots$.

\section{A Bohmian field theory in FLRW space-time}\label{flrw}

We now turn to a concrete Bohmian quantum field theory in curved space-time and point out what happens in this model with respect to the problems of cosmological vacuum fluctuations. For our purposes, 
it is sufficient to consider quantum field theory on a flat
Friedman-Lema\^itre-Robertson-Walker (FLRW) metric, since it illustrates
all technically relevant issues. 

\subsection{Definition of the theory}

The flat FLRW metric reads
\be
\dd s^2 = \dd t^2 -  a(t)^2 \delta_{ij} \dd x^i \dd x^j = a^2(\eta) \left(\dd
\eta^2 - \delta_{ij} \dd x^i \dd x^j  \right) \,,
\label{10}
\en
where $a$ is the scale factor and $\eta$ is the conformal time, defined by $ \dd
t=a \,\dd \eta$. 

Consider first a classical real scalar field $\varphi$ on this
space-time, with equation of motion given by
\be
{\ddot \varphi} + 3 \frac{{\dot a}}{a}  {\dot \varphi} - \frac{1}{a^2} \nabla^2
\varphi = 0\,,
\label{11}
\en
where the dots denote derivatives with respect to cosmic time $t$. Introducing
the field $y = a \varphi$ and using conformal time $\eta$, with $\eta$-derivatives denoted by
primes, the equation of motion reads
\be
y'' - \nabla^2 y - \frac{a''}{a} y=0 \,.
\label{12}
\en
In terms of Fourier modes, defined through
\begin{equation}
y(\eta, {\bf x})=\int{\frac{\dd^3k}{(2\pi)^{3/2}}\, y_{\bf k}(\eta) \, \ee^{\ii {\bf k}
\cdot {\bf x}}} \,,
\label{13}
\end{equation}
where $y^*_{\bf k} = y_{-{\bf k}}$ (due to the reality of $y(\eta, {\bf x})$),
this becomes
\be
y_{\bf k}'' +  \left(k^2-\frac{a''}{a}\right) y_{\bf k}=0 \,,
\label{14}
\en
with $k = |{\bf k}|$.

In the case of de Sitter space-time $a=\ee^{Ht}=-1/H\eta$ (where $\eta$ runs
from $-\infty$ to $0$). Since $a''/a = 2/\eta^2$, it will dominate $k^2$
in \eqref{14} at late times (i.e., for $\eta<0$ close to $0$), so that we have
\be
y_{\bf k}''-\frac{a''}{a} y_{\bf k} \approx 0 \,.
\label{15}
\en
Since the differential equation $y''-(2/\eta^2)y=0$ has the general solution $y(\eta) = A \, \eta^{-1} + B\, \eta^2$ with constants $A$ and $B$, we obtain that $\varphi(\eta)=y/a=-A H - B H \eta^3\approx \mathrm{const.}$ for $\eta$ near 0. This means that for large
times the field $\varphi$ becomes static. Put differently, freezing occurs in the classical theory.

Let us now turn to a quantum field on the FLRW space-time~\cite{polarski96}. In
terms of the functional Schr\"odinger picture, i.e., representing the state vector as a time-dependent functional $\Psi(y)$ of the field configuration $y({\bf x})$, the Schr\"odinger equation for
the wave functional $\Psi(y,\eta)$ is
\be\label{Psiflrw}
\ii \frac{\partial\Psi}{\partial\eta} = 
\frac{1}{2} \int \dd^3x \left[ -\frac{\delta^2}{\delta y^2} + \delta_{ij}\, \pa^i y\,
\pa^j y - \ii \frac{a'}{a}\left(\frac{\delta}{\delta y} y +
y\frac{\delta}{\delta y} \right)\right]\Psi \,.
\en
Equivalenty, in terms of Fourier modes $y_{\bf k}$, the Schr\"odinger equation
becomes
\begin{equation}\label{Psiflrwk}
\ii \frac{\partial\Psi}{\partial\eta} = 
\int_{{\mathbb R}^{3+}} \dd^3k \left[ -\frac{\delta^2}{\delta y_{\bf k}^*\delta
y_{\bf k}}+
k^2 \, y_{\bf k}^* \, y_{\bf k}
- \ii \frac{a'}{a}\left(\frac{\delta}{\delta y_{\bf k}^*}y_{\bf k}^*+
y_{\bf k}\frac{\delta}{\delta y_{\bf k}}\right)\right]\Psi
\end{equation}
for the wave functional $\Psi(y,y^*, \eta)$ (we formally treat $y_{\bf
k},y^*_{\bf k}$ as independent fields, as usual for Wirtinger derivatives; equivalently, their real and imaginary
components can be used). The integration is restricted to half the number of possible
modes (denoted by ${\mathbb R}^{3+}$), so that only independent ones are
included. 

The Bohmian approach with a field ontology can straightforwardly be applied to bosonic quantum fields on curved space time if a preferred foliation into spacelike hypersurfaces is granted, see e.g.\ \cite{hiley95,pinto-neto12a}; here we use the foliation given by the hypersurfaces $t=\mathrm{const.}$ (which corresponds to $\eta=\mathrm{const.}$); see \cite{duerr14} for a discussion of possible laws governing the foliation. In the case of the
FLRW space-time, the equation of motion for the actual field
reads
\be\label{fieldguidance}
y({\bf x})' = \frac{\delta S}{\delta y ({\bf x})} +  \frac{a'}{a}y({\bf x})
\en
or in terms of Fourier modes
\be\label{fieldguidancek}
y'_{\bf k}= \frac{\delta S}{\delta y^*_{\bf k}}+\frac{a'}{a}y_{\bf k} \,, \quad
{y^*_{\bf k}}'= \frac{\delta S}{\delta y_{\bf k}} + \frac{a'}{a}y^*_{\bf k}  \,,
\en
where $\Psi = |\Psi| \ee^{\ii S}$. 

For a product wave functional 
\be
\Psi = \prod_{{\bf k} \in {\mathbb R}^{3+}} \Psi_{\bf k}(y_{\bf k},y^*_{\bf
k},\eta) \,, 
\label{50}
\en
the functional Schr\"odinger equation reduces to the following partial
differential equation for each $\Psi_{\bf k}$,
\begin{equation}\label{Psiketa}
\ii\frac{\partial\Psi_{\bf k}}{\partial\eta}=
\left[ -\frac{\partial^2}{\partial y_{\bf k}^*\partial y_{\bf k}}+
k^2 y_{\bf k}^* y_{\bf k}
- \ii\frac{a'}{a}\left(\frac{\partial}{\partial y_{\bf k}^*}y_{\bf
k}^*+
y_{\bf k}\frac{\partial}{\partial y_{\bf k}}\right)\right]\Psi_{\bf k}\,,
\end{equation}
and the equation of motion reduces
to
\be\label{reducedfieldguidance}
y'_{\bf k}= \frac{\pa S_{\bf k}}{\pa y^*_{\bf k}}+\frac{a'}{a}y_{\bf k}
\,,
\quad {y^*_{\bf k}}'= \frac{\pa S_{\bf k}}{\pa y_{\bf k}} + \frac{a'}{a}y^*_{\bf
k}  \,,
\en
where $\Psi_{\bf k} = |\Psi_{\bf k}| \ee^{\ii S_{\bf k}}$. In quantum
equilibrium, the distribution of each mode is given by $|\Psi_{\bf k}|^2$.

\subsection{Freezing in the Bunch-Davies state}
\label{sec:flrwfreezing}

In this setting, there are several quantum states considered vacuum states. One choice, 
which is of the form of a product \eqref{50}, is given by 
\begin{equation}
\Psi_{\bf k} = \frac{1}
{\sqrt{2\pi}|f_k(\eta)|} \exp{\left\{-\frac{1}{2|f_k(\eta)|^2}|y_{\bf k}|^2 +  
\ii \left[\left(\frac{|f_k(\eta)|'}{|f_k(\eta)|}-
\frac{a'}{a}\right)|y_{\bf k}|^2-
\int_{\eta_i}^\eta \frac{\dd {\tilde \eta}}{2|f_k({\tilde
\eta})|^2}\right]\right\}} ,
\end{equation}
where $f_k$ is a solution to the classical
field equations~\eqref{14} that only
depends on the magnitude $k$ of the wave vector $\bf k$, and $\eta_i<0$ is arbitrary. 
This state is homogenous and isotropic. For this
state, the equation of motion \eqref{fieldguidance}, or \eqref{reducedfieldguidance},
is immediately integrated to yield 
\be
y_{\bf k}(\eta) = b_{\bf k}|f_k(\eta)|\,,
\en
where $b_{\bf k}$ does not depend upon $\eta$.

In the case of de Sitter space-time, it is customary to choose the Bunch-Davies
vacuum, for which 
\be
f_k = \frac{1}{\sqrt{2k}} \ee^{-\ii k\eta}\left(1 - \frac{\ii}{k\eta} \right)
\,.
\en
For early times, $\eta \to -\infty$, we have that $|f_k|\
\approx\ 1/\sqrt{2k}$, so that
the Bohmian field modes $y_{\bf k}(\eta)$ are approximately static, i.e., 
\be
y_{\bf k}(\eta) \approx c_{\bf k} \,,
\en
where $c_{\bf k}$ does not depend upon $\eta$. Thus for the original field
configuration we have 
\be
\varphi({\bf x},t)\approx \varphi({\bf x})/a
\en
in this regime. 

For large times, $\eta \to 0$, we have that $|f_k|
\approx 1/\sqrt{2k}k|\eta| =
aH/\sqrt{2k}k $, so that 
\be
y_{\bf k}(\eta) \approx d_{\bf k} a(\eta)\,,
\en
where $d_{\bf k}$ does not depend upon $\eta$.
Thus the field evolves classically in this regime. In terms of the original
field
configuration $\varphi$ and in terms of cosmological time $t$, we have that
\be\label{static}
\varphi_{\bf k}(t) \approx \varphi_{\bf k}(\infty) \quad \text{as $t\to\infty$,}
\en 
i.e., every field mode becomes static ($t$-independent) at late times---it freezes.

As a side remark, let us comment on the popular picture of the vacuum 
as a fluctuating soup of particles that spontaneously appear and
disappear. This picture arises partly, in the same way as discussed around \eqref{Xinfty} above for the hydrogen ground state, from the nonzero energy, the nonzero probability of nonzero momenta, and the thought that all configurations should be visited over time according to the $|\Psi|^2$ distribution. It also partly arises from the Feynman diagrams involving particle creation and annihilation that contribute to expressions for vacuum states. Despite all this, it is not necessary that a quantum vacuum state involves a flurry of motion, or even any motion at all, as Bohmian mechanics illustrates.

Let us be more precise about the freezing of $\varphi_{\bf k}$. As we can see from the exact time dependence,
\be\label{BDvarphit}
\varphi_{\bf k} (t) = \varphi_{\bf k}(\infty) \sqrt{1+k^2\exp(-2Ht)/H^2}\,,
\en
modes with higher $k$ freeze later (at time const.\ + $H^{-1}\log k$). Put differently, at any time $t$ those $\varphi_{\bf k}(t)$ will be close to their limiting value $\varphi_{\bf k}(\infty)$ that have $k^2\exp(-2Ht)/H^2\ll 1$ or $k\ll He^{Ht}$. Due to the expansion of de Sitter space-time, a coordinate difference $\Delta {\bf x}$ corresponds to a metric (physical) spacelike distance of $\ee^{Ht}\,\Delta {\bf x}$, so a $k$-value of $H\ee^{Ht}$ corresponds to a wave length in coordinates of $\Delta {\bf x} =H^{-1}\ee^{-Ht}$, or $1/H$ in metric distances. That is, at any time only the modes with wave lengths large compared to the Hubble distance $1/H$ are frozen.

However, the modes with smaller wave lengths, while not yet static, follow the very simple growth behavior given by \eqref{BDvarphit}.

\subsection{No Boltzmann brain problem}
\label{sec:noboltz}

For the Boltzmann brain problem, the simple behavior given by \eqref{BDvarphit} is as good as freezing, in view of the discussion in section~\ref{sec:freezingBoltzmann} above. According to the ``optimistic'' view described there, a Boltzmann brain needs to be functioning, which requires complex dynamical behavior incompatible with \eqref{BDvarphit}. According to the ``pessimistic'' view, configurational structures encoding memories or thoughts are relevant, and mere growth according to the factor $\sqrt{1+k^2\exp(-2Ht)/H^2}$ would not change the structure in a relevant way---it would not change what, if anything, is encoded in the configuration $\varphi$. If at some late time $t_0$, with probability close to 1, there are no brain configurations (encoding memories or thoughts) in $\varphi$, then no such configurations arise from \eqref{BDvarphit} at any later time. Thus, the problem associated with Boltzmann brains is absent in the Bunch-Davies state.

Now the quantum state $\Psi$ at late times will not be exactly the Bunch-Davies state. In fact, it will not even be close in Hilbert space to the Bunch-Davies state; it looks like the Bunch-Davies state for relevant local observables, but the Bohmian motion depends on the wave function in a non-local way. So the question arises whether freezing, or simple growth as in \eqref{BDvarphit}, also occurs for other wave functions than the Bunch-Davies state. 

This question is answered in \cite{tumulka15}, where it is derived that, in the Bohmian model defined by \eqref{Psiflrw} and \eqref{fieldguidance}, the time dependence \eqref{BDvarphit} applies asymptotically at late times to \emph{every} Bohmian history $\varphi$ and \emph{generic} wave function $\Psi$. In fact, the result is that \eqref{BDvarphit} holds to arbitrary degree of accuracy, simultaneously for all $\bf k$, for sufficiently large $t$. As above for the Bunch-Davies state, we conclude for generic quantum states that Boltzmann brains do not occur. Thus, this Bohmian model is free of a Boltzmann brain problem, and we surmise that this is so for natural Bohmian models in general.

\section{Bohmian fluctuations in inflation theory} 
\label{bohmfluctuations}

We have described in section~\ref{sec:breaking} how Bohm-type theories can in principle account for inhomogeneity in the density of matter even when the quantum state is homogeneous and isotropic. Now we outline how this works in a concrete model.

\subsection{Bohmian dynamics during inflation}
\label{sec:Bohminflation}

In inflation theory, one needs to consider a quantized metric along with the inflaton field and ordinary matter, so the model of section~\ref{flrw} does not suffice. However, the full model can be simplified as follows. One considers small perturbations of the inflaton field and the metric around a ``background'' solution of the classical equations for the inflaton field and the metric; the linearized equations for these perturbations are then quantized. In this quantum theory, the crucial role is played by the gauge invariant Mukhanov-Sasaki variable, a linear combination of the (linear perturbation of the) quantum field and the gravitational potential (the quantum field representing the linearized perturbations in the metric). In a Bohmian version of the theory with a field ontology, the (operator-valued) quantum fields and quantized metric are supplemented with an actual (real-valued) field configuration and an actual metric. As a consequence, also the (operator-valued) Mukhanov-Sasaki variable is supplemented with an actual (real-valued) configuration. 
It is the latter that provides the inhomogeneity needed for the seeds of structure formation, breaking the symmetry of the state vector alone.

The equations governing the Mukhanov-Sasaki variable are actually very similar to Equations~\eqref{Psiflrw}--\eqref{fieldguidancek} in section~\ref{flrw}; one only needs to replace $a$ by $z= a^2 \phi'_0/a'$, with $\phi_0$ the background scalar field, and interpret $y$ as the Mukhanov-Sasaki variable. The time evolution of $a$ depends on the inflationary model. An analysis analogous to that of section \ref{flrw} shows that  at late times, i.e., at the end of the inflationary period, the Bohmian (real-valued) $y$ behaves classically and, not surprisingly, features inhomogeneities. 
The details are given in \cite{pinto-neto12a}.  (In \cite{hiley95}, a similar analysis is given, but
instead of using the Mukhanov-Sasaki variable, only the fluctuations of the
inflaton field are treated as quantum. The metric perturbation then is
derived from the inflaton perturbations using classical relations.)  

In the inflation literature it is often argued (e.g., \cite{guth85,albrecht94,polarski96}) that the quantum state of the universe ``becomes classical'' at the end of the inflationary period. Various arguments are given: the validity of the WKB approximation, the resemblance of the Wigner distribution to a classical phase-space distribution, etc.. These seem to suggest that the wave function  somehow represents a classical ensemble of trajectories. In Bohmian mechanics, this makes clear sense: Every wave function is associated with an ensemble of trajectories, and we can ask whether most of these trajectories are approximately classical (i.e., whether they satisfy classical equations). In orthodox quantum mechanics, however, where the existence of trajectories is denied, the meaning of ``becoming classical'' is rather obscure. If both the position and momentum uncertainty of a wave packet are simultaneously rather small (within the limits of the Heisenberg uncertainty relation) then it can be meaningful to talk about whether the wave packet as a whole moves along a classical trajectory; but the wave function relevant to inflation is, in fact, very spread out in field configuration space. Or, one might want to say that a quantum system behaves classically if observations of it are consistent with a classical model; but, in the situation of interest here, there are no observers around to perform observations. 

In the Bohmian model, as noted above, the trajectories do indeed become classical at the end of the inflationary period, as desirable in inflation theory. On the suggestion \cite{polarski96} that decoherence arising from coupling to other degrees of freedom helps with obtaining classical motion, we have two remarks. First, in orthodox quantum mechanics, decoherence alone does not help in the absence of observers, for the reasons described in the previous paragraph, just as decoherence alone does not solve the quantum measurement problem. Second, in Bohmian mechanics, as we just saw, the trajectories become classical in this situation regardless of whether decoherence occurs.

Two toy examples from non-relativistic quantum mechanics often discussed in the context of inflation theory to illustrate the emergence of classicality are the Mott problem and the upside-down harmonic oscillator. For the Mott problem, we have discussed in section~\ref{sec:breaking} how classical motion emerges from Bohmian mechanics, while orthodox quantum mechanics does not provide a full and clear account of such motion. The situation is similar for the upside-down harmonic oscillator, i.e., a single particle in 1 dimension with Hamiltonian $\hat{H}= -\tfrac{1}{2} \partial_x^2-\tfrac{1}{2}x^2$: The Bohmian motion approaches Newtonian motion for late times, while the wave function tends to become spread-out in both position and momentum space, so that orthodox quantum mechanics lacks clear justification for regarding it as ``classical.''

\subsection{Probability and observation}

We see only one sky, which is produced by one actual configuration of the fields. Yet
quantum theory gives us statistical predictions. How should we understand the
statistical predictions? 

To answer this question, let us focus on the cosmic microwave background (CMB) anisotropies. Let $T({\bf n})$ denote the temperature of the CMB in the direction ${\bf n}$, with ${\bar T}$ its average over the sky. The temperature anisotropy $\delta T({\bf n}) / {\bar T}$, where $\delta T({\bf n}) = T({\bf n}) - {\bar T}$, can be expanded in terms of spherical harmonics 
\be
\frac{\delta T({\bf n})}{{\bar T}} = \sum^{\infty}_{l=2} \sum^{m=l}_{m=-l}
a_{lm} Y_{lm}({\bf n}) \,.
\en
The $a_{lm}$ are determined by the Mukhanov-Sasaki variable. The main quantity
used to study the temperature anisotropies is the angular power spectrum
\be
C^0_l = \frac{1}{2l+1} \sum_m |a_{lm}|^2 \,.
\en
In the standard treatments, one considers the operator ${\widehat C^0_l}$ and
compares the observed value for the angular power spectrum with $C_l =
\langle \Psi | {\widehat C^0_l} |\Psi \rangle$. It is sometimes claimed that
this expectation value, which corresponds to the average over an ensemble of
universes, will agree with an average of the angular power spectrum seen for
different observers at large spatial separations. While this may be true,
it does not seem relevant, since we do not have
observations from other places, just from earth. 

What is relevant, however, is the variance. Suppose for simplicity that 
the Mukhanov-Sasaki variables are jointly Gaussian (this is the case in the simplest inflationary models, but most inflationary models predict deviations). Then also 
the $a_{lm}$ are jointly Gaussian, and the variance of $C^0_l$ is given by \cite{lyth09}
\be
(\Delta C^0_l)^2 = \frac{2}{2l+1}  C_l^2 \,.
\en
This means that one would expect the observed value to deviate from $C_l$ by an
amount of the order $\sqrt{2/(2l+1)}C_l$. We will hence have a greater uncertainty for
small $l$ (which corresponds to large angles over the sky, since the angle is
roughly $\pi/l $). For large $l$ the observed value must lie closer to the
expected value.

This kind of reasoning is justified in the Bohmian approach (while in the orthodox interpretation of quantum theory there remains a gap, viz., the measurement problem). Indeed, since in Bohmian mechanics the initial configuration $Q_0$ of the universe is a realization of (i.e., typical of) the $|\Psi_0|^2$ distribution, the $a_{lm}$ obtained from $Q_0$ are a realization of the joint distribution that follows from $|\Psi_0|^2$ which, as we assumed, is Gaussian in the case at hand. And for a realization of a Gaussian random variable, its deviation from the expectation value is indeed of the order of the root mean square.

\section{Concluding Remark}

The absence of an observer external to the universe is awkward for orthodox quantum cosmology. Without such an observer one wonders why the usual measurement rules for the appearance of quantum probabilities should be relevant. For Bohmian mechanics, however, there is no such awkwardness: the desirable fluctuations (inhomogeneities in the matter density in the early universe) do occur also in the absence of an external observer (or wave function collapse), and the undesirable fluctuations (Boltzmann brains in the late universe) presumably do not occur, exactly because there are no external observers causing the wave function to collapse.

\section{Acknowledgments}
All three authors acknowledge support from the John Templeton Foundation, grant no.~37433.
W.S.\ acknowledges support from the Actions de Recherches Concert\'ees (ARC) of
the Belgium Wallonia-Brussels Federation under contract No.\ 12-17/02.

\bibliography{ref.bib}

\end{document}